\newcommand{\msbar}{{\overline {\rm MS}}}
\def\lsim{\raise0.3ex\hbox{$<$\kern-0.75em\raise-1.1ex\hbox{$\sim$}}}
\def\gsim{\raise0.3ex\hbox{$>$\kern-0.75em\raise-1.1ex\hbox{$\sim$}}}
\def\simgt{\rlap{\lower 4.0 pt\hbox{$\mathchar \sim$}}\raise 1.0pt \hbox {$>$}}
\def\simlt{\rlap{\lower 4.0 pt\hbox{$\mathchar \sim$}}\raise 1.0pt \hbox {$<$}}
\documentclass{PoS}

\title{Charm quark system in $2+1$ flavor lattice QCD using the PACS-CS configurations}

\ShortTitle{Charm quark system in $2+1$ flavor lattice QCD using the PACS-CS configurations}

\author{PACS-CS Collaboration : 
\speaker{Y. Namekawa}${}^{a}$\thanks{E-mail: namekawa@ccs.tsukuba.ac.jp},
 S.~Aoki${}^{b,c}$,
 N. Ishii${}^{a}$,
 K.-I.~Ishikawa${}^{d}$,
 N.~Ishizuka${}^{a,b}$,
 T. Izubuchi${}^{c,e}$,
 D. Kadoh${}^{a}$,
 K.~Kanaya${}^{b}$,
Y. Kuramashi${}^{a,b}$,
 M.~Okawa${}^{d}$,
 Y.~Taniguchi${}^{a,b}$,
 A.~Ukawa${}^{a,b}$,
 N.~Ukita${}^{a}$,
 T.~Yoshi\'e${}^{a,b}$
 \\
 \llap{${}^a$}Center for Computational Sciences, University of Tsukuba, Tsukuba, Ibaraki 305-8577, Japan\\
 \llap{${}^b$}Graduate School of Pure and Applied Sciences, University of Tsukuba, Tsukuba, Ibaraki 305-8571, Japan\\
 \llap{${}^c$}Riken BNL Research Center, Brook-haven National Laboratory, Upton, New York 11973, USA\\
 \llap{${}^d$}Graduate School of Sciences, Hiroshima University, Higashi-Hiroshima, Hiroshima 739-8526, Japan\\
 \llap{${}^e$}Institute for Theoretical Physics, Kanazawa University, Kanazawa, Ishikawa 920-1192, Japan}

\abstract{
We study heavy-heavy and heavy-light quark systems for charm 
with a relativistic heavy quark action in $2+1$ flavor
lattice QCD. Configurations are generated by the PACS-CS
Collaboration at the lattice spacing is $a=0.09$ fm
with the lattice size of $32^3\times 64$ employing the
$O(a)$-improved Wilson quark action and the Iwasaki gauge action. 
We present preliminary results for the charmonium spectrum and
the $D$ and $D_s$ meson decay constants evaluated at 3.5 MeV$< m_{\rm ud}<$
12 MeV with $m_{\rm s}$  around the physical value. 
We investigate the dynamical quark mass dependences of the hyperfine
and the orbital splittings. The decay constants are compared with 
the recent experimental values. 
}

\FullConference{The XXVI International Symposium on Lattice Field Theory\\
		July 14-19 2008\\
		Williamsburg, Virginia, USA}

\begin{document}

\section{Introduction}
\label{section:introduction}

Precise determination of physical quantities for heavy quark systems 
provides us with an opportunity to search for new physics beyond the
standard model. For this purpose lattice QCD should be a powerful tool. 
However, the use of conventional lattice quark actions are problematic 
because of large cutoff errors due to the heavy quark masses.
So far several approaches to avoid this problem 
have been employed for the study of heavy 
quark physics on the lattice\cite{Gamiz}. 
Our choice is to employ the relativistic heavy quark action of 
Ref.~\cite{akt}.  This formalism 
allows us to take the continuum limit in which $m_Q a$
corrections are controlled by a smooth function. 
In fact the cutoff errors are reduced
from $O((m_Q a)^n)$
to $O(f(m_Q a)(a \Lambda_{QCD})^2)$
where $f(m_Q a)$ is an analytic function around $m_Q a = 0$.

\section{Simulation parameters}
\label{sec:param}

We simulate the charm quark system with the relativistic
heavy quark action of Ref.~\cite{akt} on the 2+1 flavor lattice QCD
configurations which are generated by the PACS-CS Collaboration
employing the nonperturbatively $O(a)$-improved Wilson quark action 
with $c_{\rm SW}^{\rm NP}=1.715$\cite{csw_np} and the Iwasaki gauge action.
The lattice size is $32^3\times 64$ whose spatial extent is $L=2.9$ fm
with the lattice spacing of $a=0.09$ fm. The dynamical up-down quark mass 
ranges from 67 MeV down to 3.5 MeV which is close to the 
the physical value. The details of the configuration production and light 
hadron physics that emerges from them are described 
in Refs.~\cite{pacscs,kura_lat08,ukita_lat08,kadoh_lat08}.
Table~\ref{tab:stat} summarizes the simulation parameters and the
statistics of the configuration sets we have used for the heavy quark 
measurements.
The number of the source points is to be quadrupled.
We emphasize that the data point at 
($\kappa_{\rm ud},\kappa_{\rm s}$)=(0.137785,0.13660) is almost
on the physical point: the up-down quark mass is only 30\% heavier 
and the strange quark mass is almost exactly at the physical value.

\begin{figure}[t]
\begin{center}
 \includegraphics[width=75mm]{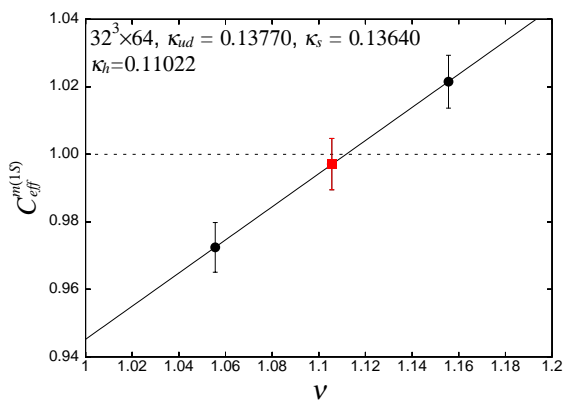}
 \caption{
  Nonperturbative tuning of $\nu$ with $\kappa_{\rm h}=0.11022$
at ($\kappa_{\rm ud},\kappa_{\rm s}$)=(0.13770,0.13640). A red symbol
 denotes the result for the nonperturbative $\nu$.
 }
 \label{fig:nu_np}
\end{center}
\end{figure}

The relativistic heavy quark action
proposed in Ref.~\cite{akt} is given by
\begin{eqnarray}
 S_Q & = & \sum_{x,y}\overline{Q}_x D_{x,y} Q_y,\\
D_{x,y} &=& \delta_{xy}
- \kappa_{\rm h}
  \sum_i \left[  (r_s - \nu \gamma_i)U_{x,i} \delta_{x+\hat{i},y}
                +(r_s + \nu \gamma_i)U_{x,i}^{\dag} \delta_{x,y+\hat{i}}
         \right]
 \nonumber \\
&&- \kappa_{\rm h}
         \left[  (r_t - \nu \gamma_i)U_{x,4} \delta_{x+\hat{4},y}
                +(r_t + \nu \gamma_i)U_{x,4}^{\dag} \delta_{x,y+\hat{4}}
         \right]
 \nonumber \\
&&- \kappa_{\rm h}
  \left[
     c_B \sum_{i,j} F_{ij}(x) \sigma_{ij}
  +  c_E \sum_i     F_{i4}(x) \sigma_{i4}
  \right],
\end{eqnarray}
where we are allowed to choose $r_t=1$, while the other
four parameters $\nu$, $r_s$, $c_B$, $c_E$ should be adjusted
in the mass dependent way.
We use the one-loop perturbative values for $r_s$, $c_B$ and $c_E$ 
evaluated in Ref.~\cite{param}.
For the clover coefficients $c_{B}$ and  $c_{E}$
we incorporate the nonperturbative contributions at the massless limit
adopting the procedure 
$c_{B,E}=(c_{B,E}(m_Q a) - c_{B,E}(0))^{\rm PT} + c_{\rm SW}^{\rm NP}$.
At each simulation point,
the parameter $\nu$ is nonperturbatively determined
from the dispersion relation for the spin-averaged $1S$ state of the charmonium:
\begin{equation}
   E({\vec p})^2
 = E({\vec p})^2+c_{\rm eff}^2 |{\vec p}|^2,
\end{equation}
where $\nu$ is adjusted such that the effective speed of light $c_{\rm eff}$
becomes unity.
Figure~\ref{fig:nu_np} shows an example
of the nonperturbative tuning of $\nu$ with $\kappa_{\rm h}=0.11022$
at ($\kappa_{\rm ud},\kappa_{\rm s}$)=(0.13770,0.13640).
In order to search for the physical charm quark mass point,
we employ two values of the hopping parameter of a heavy quark
$\kappa_{\rm h}$.
The values of $\kappa_{\rm h}$ are chosen to
sandwich the physical charm quark mass.

\begin{table}[t]
\caption{Simulation parameters. 
Quark masses are perturbatively renormalized
in the $\msbar$ scheme. 
The renormalization scale is $\mu=1/a$ for $\kappa_{\rm ud}\le 0.137785$
and $\mu=2$ GeV for the physical point.}
\label{tab:stat}
\begin{center}
\begin{tabular}{ccccccc} 
\hline
\multicolumn{1}{c}{$\kappa_{\rm ud}$} & \multicolumn{1}{c}{$\kappa_{\rm s}$} &
 \multicolumn{1}{c}{$m_{\rm ud}^{\msbar}(\mu)$} [MeV] & 
 \multicolumn{1}{c}{$m_{\rm s}^{\msbar}(\mu)$}  [MeV] & 
 \multicolumn{2}{c}{\#conf} & \#source \\
& & &  &
 measured & total/MD time &
\\ \hline
 0.13770        & 0.13640       &
 12.3(2) & 90(1) &
 400 & 800/2000 & 1
\\ 
 0.13781        & 0.13640   &
 3.5(2) & 87(1) &
 100 & 198/990  & 1
\\ 
 0.137785       & 0.13660       &
 3.5(1) & 73(1) &
 90 & 200/1000   & 1
\\ \hline
\multicolumn{2}{c}{physical point} & 2.53(5) & 72.7(8) & & & \\
\hline
\end{tabular}
\end{center}
\end{table}

\section{Heavy-heavy system}
\label{section:heavy_heavy}

\begin{figure}[t]
\begin{center}
 \includegraphics[width=75mm]{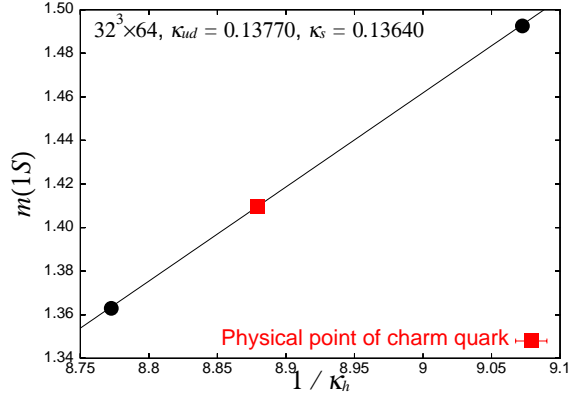}
 \caption{
  Interpolation of the spin-averaged $1S$ charmonium mass
 to the physical point as a function of $\kappa_{\rm h}$. Errors are
 within symbols.}
 \label{figure:kappa_heavy_inv-m_PS_V_nu2}
\end{center}
\end{figure}

Let us first investigate the charmonium spectrum.
At each combination of ($\kappa_{\rm ud},\kappa_{\rm s}$)
we determine the physical point of the charm quark by the condition that 
the mass of the spin-averaged 
$1S$ state reproduces the experimental value,
$M(1S) = (M_{\eta_c} + 3 M_{J/\psi})/4
 = 3.0677(3)$ [GeV]\cite{PDG}.
For this purpose we linearly interpolate the results for $M(1S)$ 
at two values of the hopping parameter $\kappa_{\rm h}$.
Figure~\ref{figure:kappa_heavy_inv-m_PS_V_nu2} illustrates
this procedure for the case of 
($\kappa_{\rm ud},\kappa_{\rm s}$)=(0.13770,0.13640).

\begin{figure}[t]
\begin{center}
 \includegraphics[width=75mm]{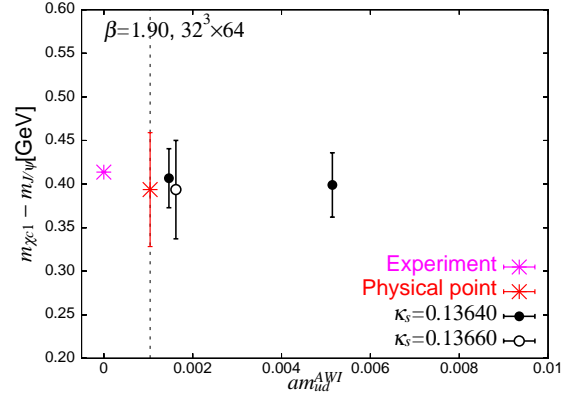}
 \caption{
  Orbital excitation $m_{\chi_{c1}}-m_{J/\psi}$ as 
a function of $m_{\rm ud}^{AWI}$. Vertical dotted line denotes the
 physical point.}
 \label{figure:bare_m_ud_AWI-m_3_P_1_minus_m_V}
\end{center}
\end{figure}

In Fig.~\ref{figure:bare_m_ud_AWI-m_3_P_1_minus_m_V}
we plot results for the mass of the orbital excitation 
$m_{\chi_{c1}}(1P) - m_{J/\psi}(1S)$ at the
physical point of the charm quark mass. It is hard to detect the
dynamical quark mass dependence within the errors
in the range of $3.5 {\rm MeV}\simlt m_{\rm ud}\simlt 12{\rm MeV}$ and 
$73 {\rm MeV}\simlt m_{\rm s}\simlt 90{\rm MeV}$.
A very short chiral extrapolation is made employing a linear function
of the up-down and the strange quark masses:
\begin{equation}
 m_{\chi_{c1}}(1P) - m_{J/\psi}(1S) = \alpha + \beta m_{\rm ud} 
+ \gamma m_{\rm s}.
\end{equation}
In Fig.~\ref{figure:bare_m_ud_AWI-m_3_P_1_minus_m_V} we find that 
the extrapolated value and its error are almost identical to  
the result at ($\kappa_{\rm ud},\kappa_{\rm s}$)=(0.137785,0.13640). 
This illustrates how close our simulation points are to the 
physical point.  The result at the physical point
is consistent with the experimental value within the error.

\begin{figure}[t]
\begin{center}
 \includegraphics[width=75mm]{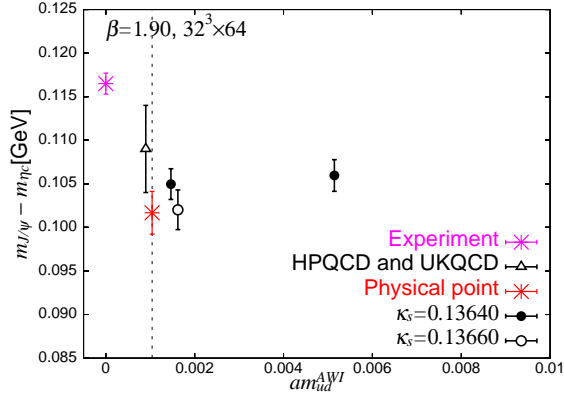}
 \caption{
  Hyperfine splitting of the charmonium as a function of $m_{\rm ud}^{AWI}$. 
 A vertical dotted line denotes the physical point.}
 \label{figure:bare_m_ud_AWI-m_V_minus_m_PS}
\end{center}
\end{figure}

Figure~\ref{figure:bare_m_ud_AWI-m_V_minus_m_PS}
shows the results for the hyperfine splitting $m_{J/\psi}-m_{\eta_c}$.
As in the orbital excitation we find little dynamical quark mass dependence.
We extrapolate the results to the physical point 
employing a linear function of the dynamical quark masses.
The extrapolated value, which is essentially determined  by the
result at ($\kappa_{\rm ud},\kappa_{\rm s}$)=(0.137785,0.13640), 
shows a 10\% deficit from the experimental value.
Possible sources of the discrepancy 
are $O(g^2 a)$ effects
in the relativistic heavy quark action, dynamical charm quark
effects and disconnected loop contributions.
A recent 2+1 flavor lattice QCD calculation 
with the highly improved staggered quarks\cite{hfs_hpqcd}
shows a similar value to ours.

\begin{figure}[t]
\begin{center}
 \includegraphics[width=75mm]{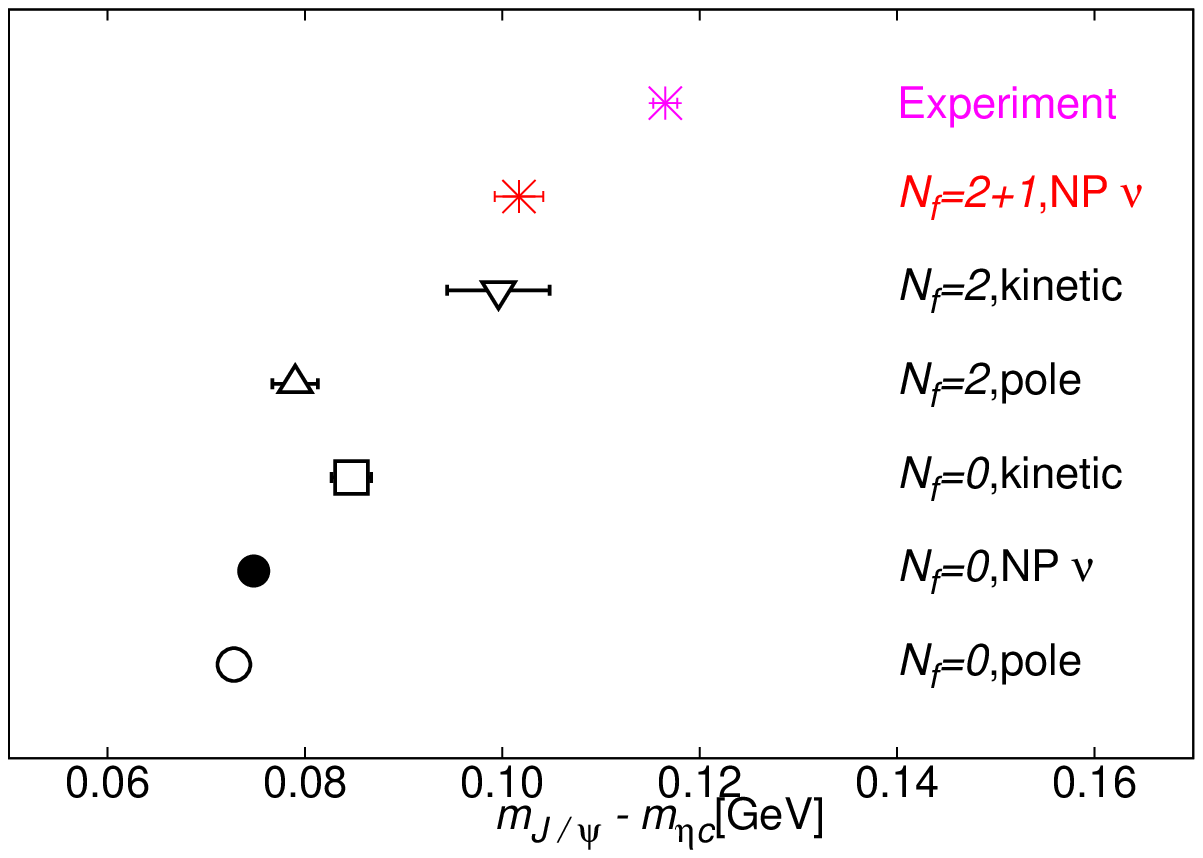}
 \caption{
Comparison of hyperfine splittings 
in $N_f=0$\cite{kura_lat05}, 2\cite{RHQ-N_f_0_2} and $2+1$,
together with the experimental value. All the lattice results are
 obtained at $a^{-1}\approx 2$ GeV.}
 \label{figure:comparison_of_m_V_minus_m_PS}
\end{center}
\end{figure}

We compare our results for the hyperfine splitting in $N_f=2+1$
QCD with the previous $N_f=0,2$ results\cite{kura_lat05,RHQ-N_f_0_2} in
Fig.~\ref{figure:comparison_of_m_V_minus_m_PS}.  We observe a clear 
trend that the results become closer to the experimental value
as the number of the flavor is increased. 
The dynamical quarks give significant contributions 
to the hyperfine splitting.

\section{Heavy-light system}
\label{section:heavy_light}

\begin{figure}[t]
\begin{center}
 \includegraphics[width=75mm]{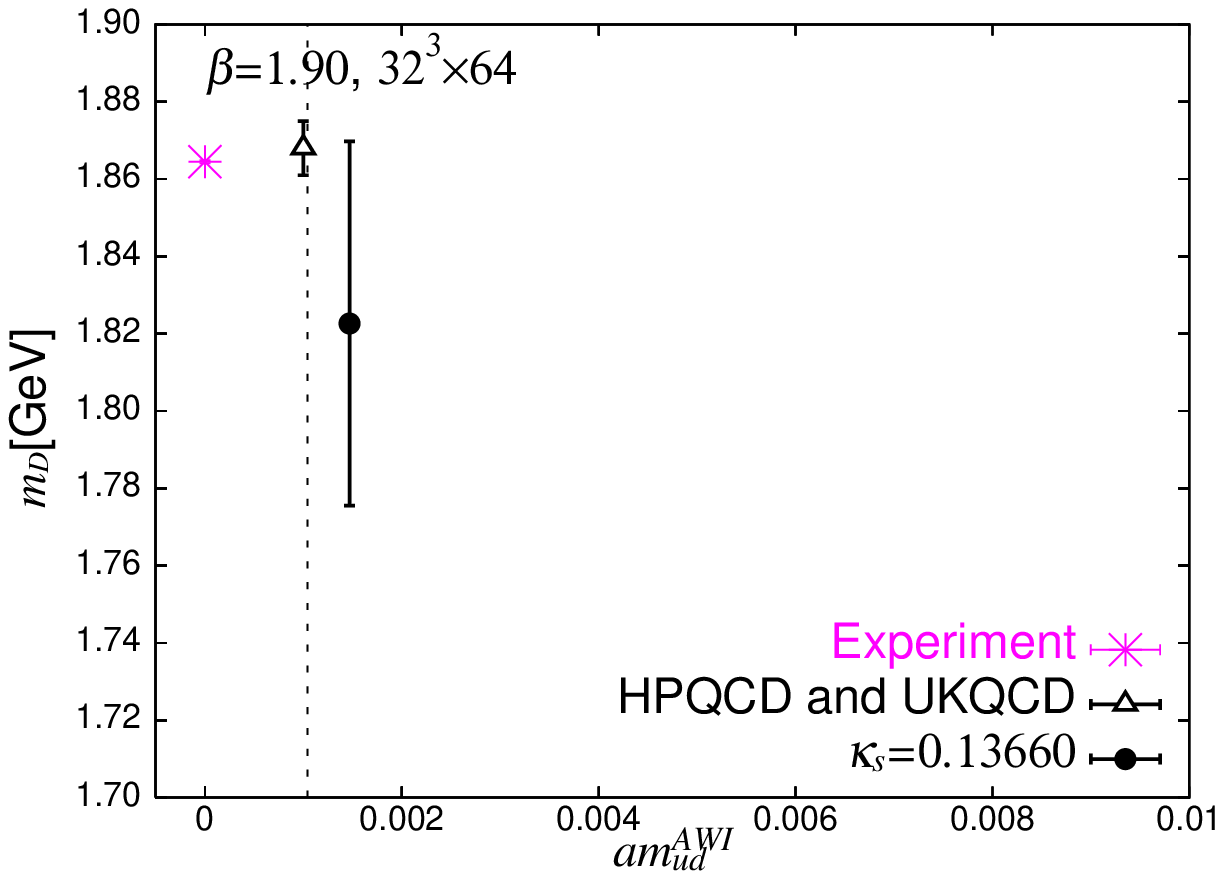}
 \includegraphics[width=75mm]{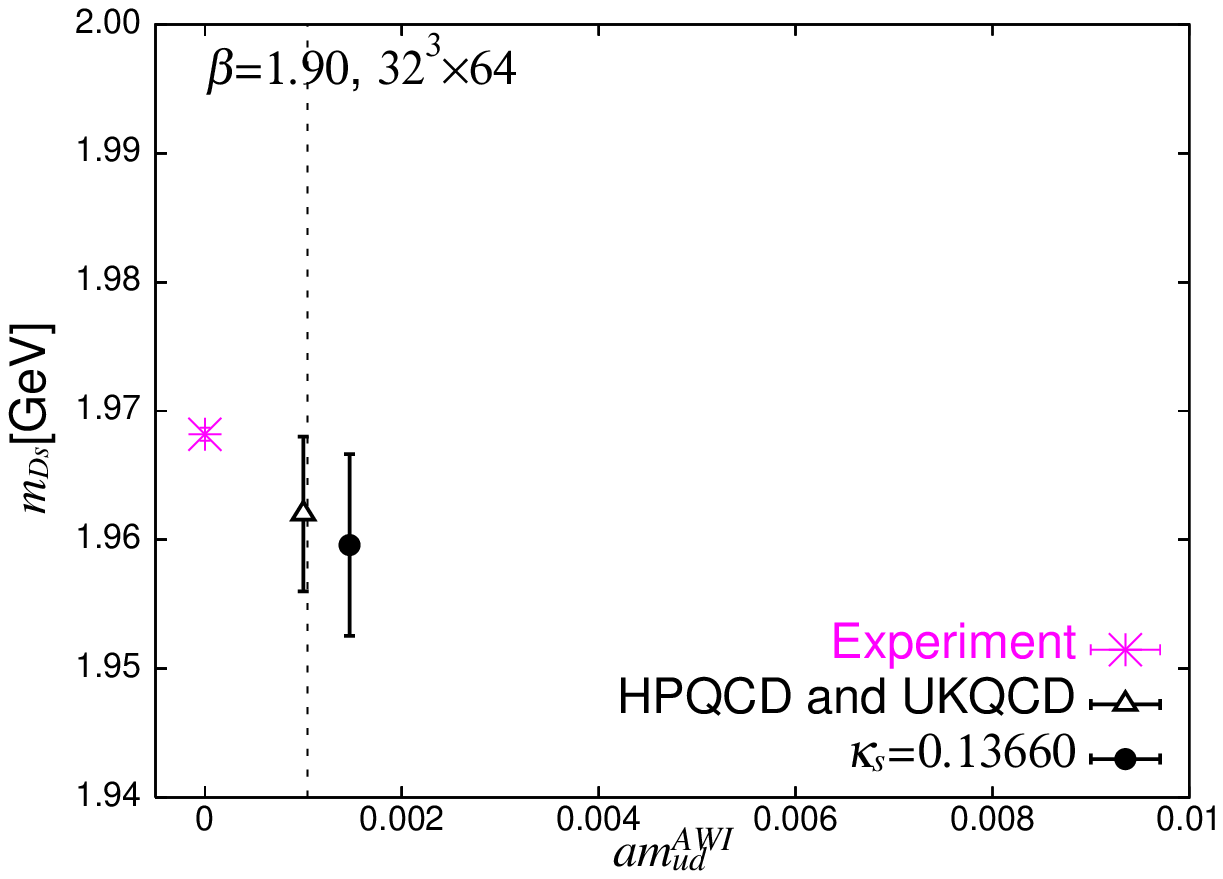}
 \caption{
  $D$ (left) and $D_s$ (right) meson masses 
as a function of $m_{\rm ud}^{\rm AWI}$. 
A vertical dotted line denotes the physical point.}
 \label{fig:mps_hl}
\end{center}
\end{figure}

\begin{figure}[t]
\begin{center}
 \includegraphics[width=75mm]{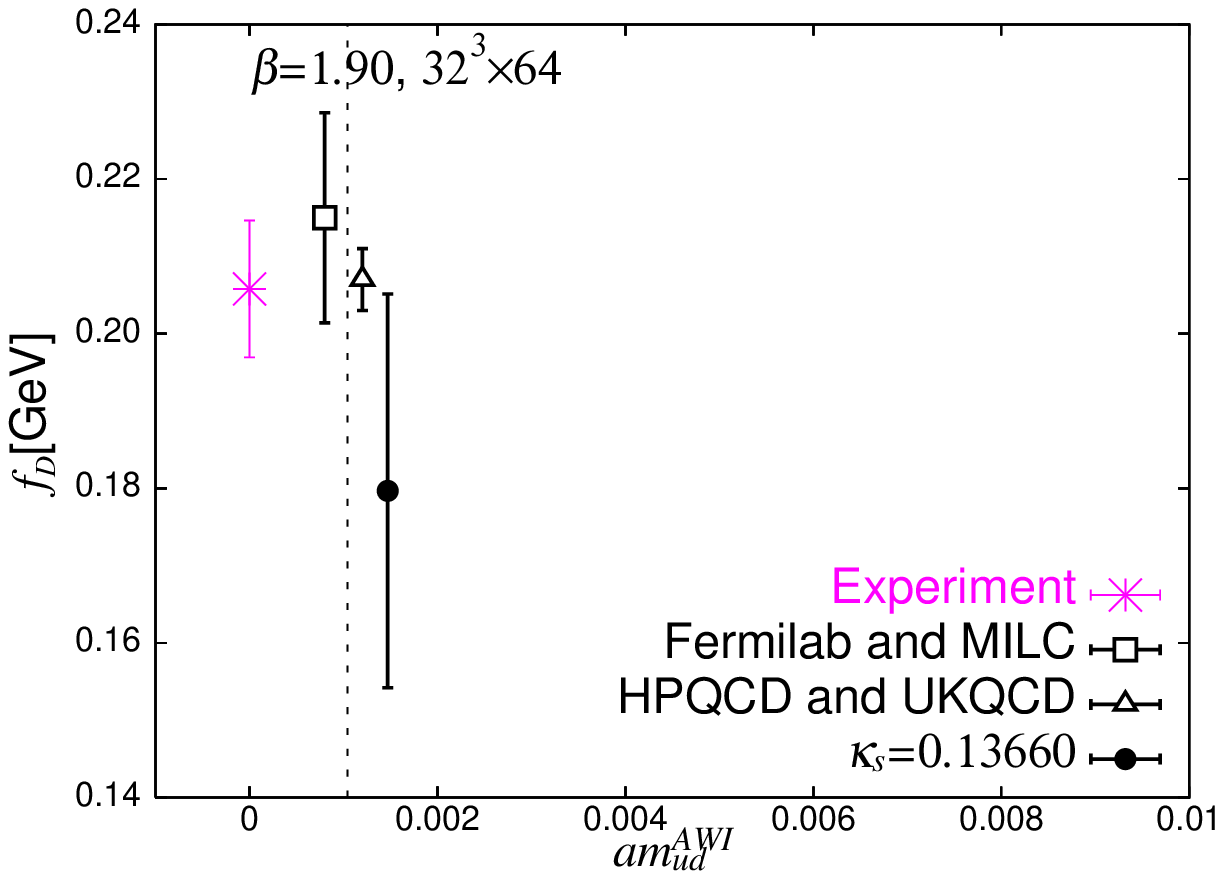}
 \includegraphics[width=75mm]{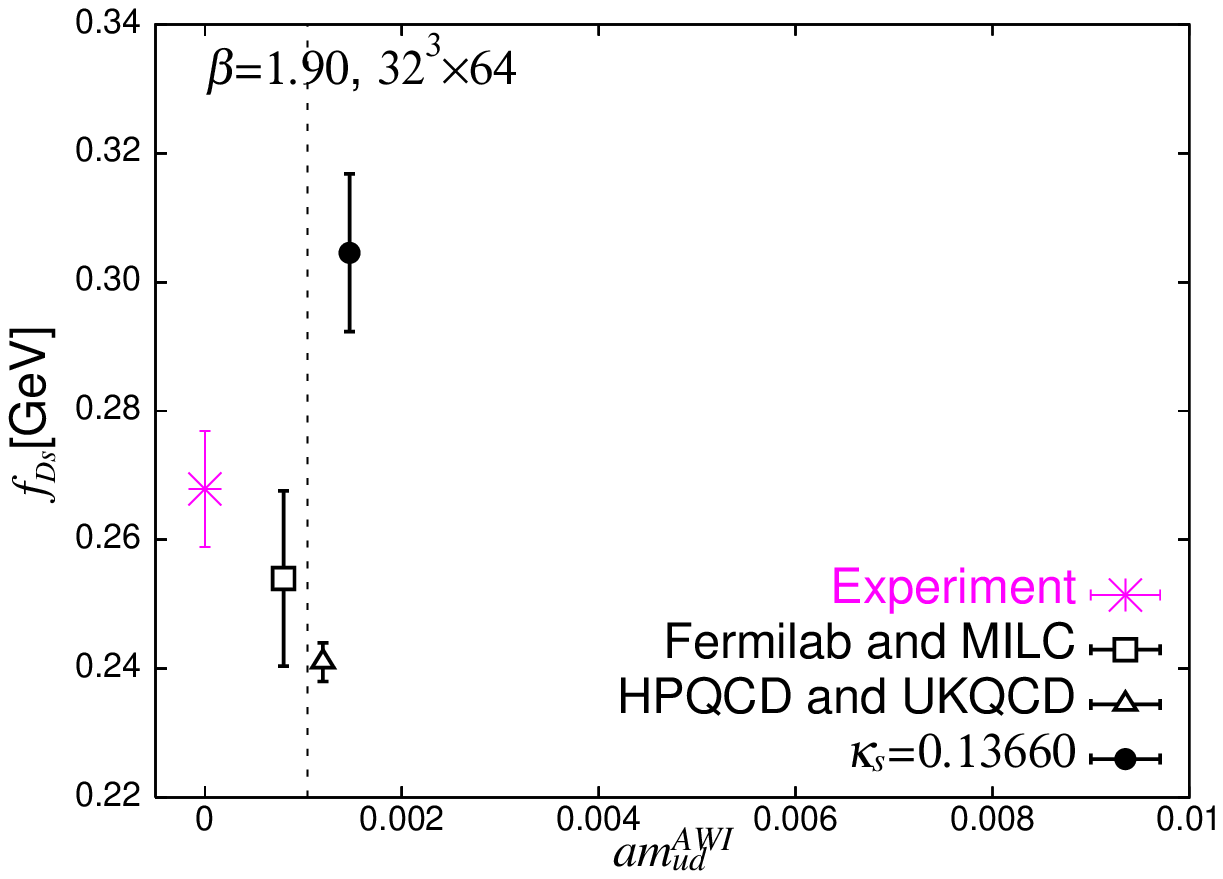}
 \caption{$f_D$ (left) and $f_{D_s}$ (right) meson masses 
as a function of $m_{\rm ud}^{\rm AWI}$. 
A vertical dotted line denotes the physical point.}
 \label{fig:fps_hl}
\end{center}
\end{figure}

For the heavy-light system
we focus on the $D$ and $D_s$ mesons and their decay constants
measured at ($\kappa_{\rm ud},\kappa_{\rm s}$)=(0.137785,0.13660).
This data point is so close to the physical point that 
$m_{\rm ud}$ corrections in the results could be
smaller than the statistical errors.
We employ perturbative values for the renormalization factor and the 
improvement coefficients of the axial vector current
evaluated in Ref.~\cite{zvza}. For $c_{A_4}^{+}$ we incorporate 
the nonperturbative contribution at the massless limit by
$c_{A_4}^{+}=(c_{A_4}^{+}(m_Q a) - c_{A_4}^{+}(0))^{\rm PT} 
+ c_{A}^{\rm NP}$ with $c_{A}^{\rm NP} = -0.03876106$\cite{NP-c_A}.
Figure~\ref{fig:mps_hl} compares our results for 
the $D$ and $D_s$ meson masses with the experimental values~\cite{CLEO}.
They are consistent within the errors.
It is noteworthy that the physical charm quark mass determined from 
the heavy-heavy system successfully reproduces the heavy-light meson masses.
The results for the decay constants are shown in Fig.~\ref{fig:fps_hl},
where we also plot the recent 2+1 flavor lattice QCD results
with relativistic heavy quark actions\cite{hpqcd,fnal} for comparison.    
Although we find a sizable discrepancy between our result
and the experimental value for $f_{D_s}$, we should analyze the full 
configuration set and improve statistics before  we derive any conclusions.    

\begin{figure}[t]
\begin{center}
 \includegraphics[width=75mm]{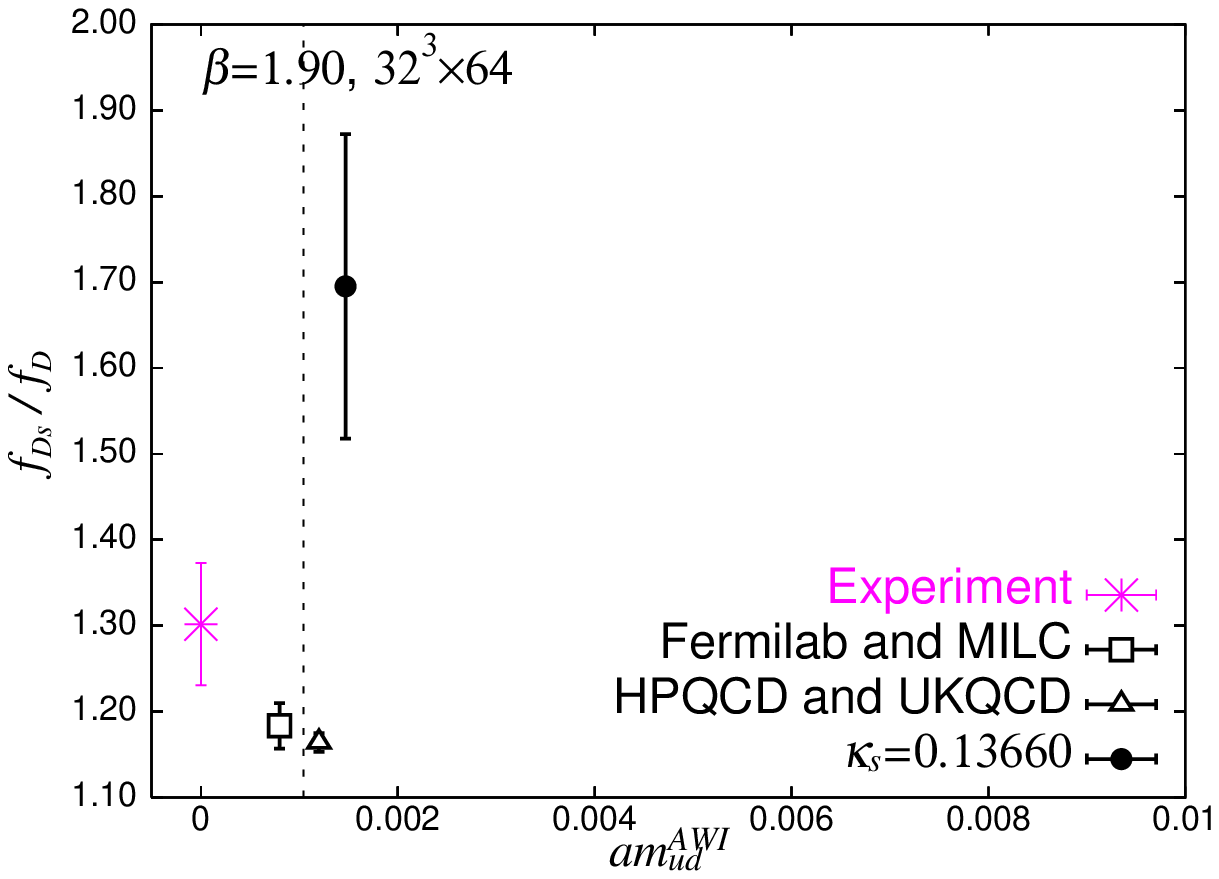}
 \includegraphics[width=75mm]{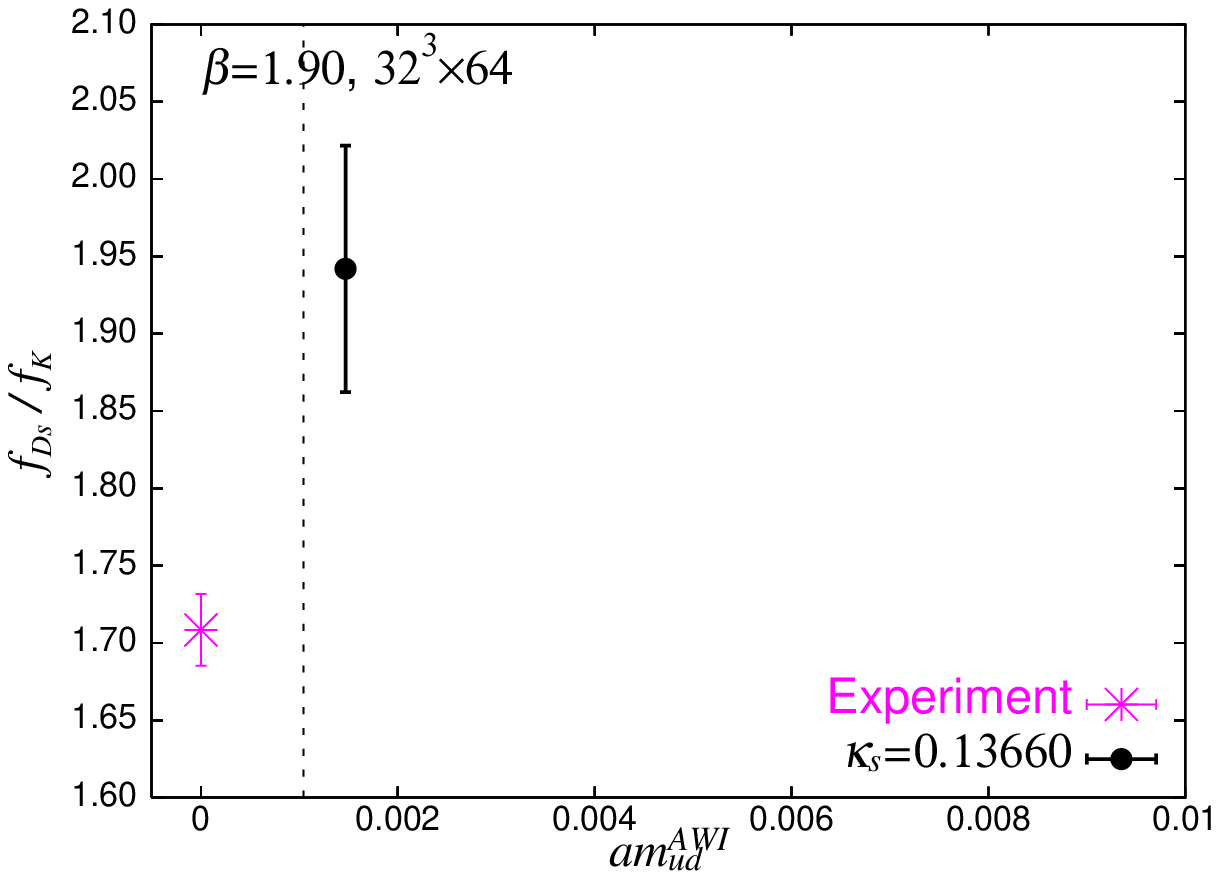}
 \caption{
  Ratios of $f_{D_s}$ to $f_{D}$ (left) and $f_{D_s}$ to $f_K$ (right).
 A vertical dotted line denotes the physical point.
 }
 \label{figure:bare_m_ud_AWI-f_D_s_over_f_K}
\end{center}
\end{figure}

In Fig.~\ref{figure:bare_m_ud_AWI-f_D_s_over_f_K} we plot
the ratios of $f_{D_s}$ to $f_D$ and $f_{D_s}$ to $f_K$ 
in which uncertainties coming from the perturbative renormalization factors 
and the lattice cutoff should cancel out. 
For both cases our results show larger values than
the experimental ones, which originate from the discrepancy
found in Fig.~\ref{fig:fps_hl}.

\begin{acknowledgments}
Numerical calculations for the present work have been carried out
on the PACS-CS computer
under the ``Interdisciplinary Computational Science Program'' of
Center for Computational Sciences, University of Tsukuba.
This work is supported in part by Grants-in-Aid for Scientific Research
from the Ministry of Education, Culture, Sports, Science and Technology
(Nos.
16740147,
17340066,
18104005,
18540250,
18740130,
19740134,
20340047,
20540248,
20740123,
20740139
).
\end{acknowledgments}


\begin{thebibliography}{99}

\bibitem{Gamiz}
For a review, see
E.~Gamiz,
\pos{PoS(LATTICE 2008)014}.


\bibitem{akt} 
S.~Aoki, Y.~Kuramashi and S.~Tominaga,
{\em Prog. Theor. Phys.} {\bf 109} (2003) 383.

\bibitem{csw_np}
CP-PACS and JLQCD Collaborations, S. Aoki {\it et al.},
{\em Phys. Rev.} {\bf D73}, 034501 (2006).



\bibitem{pacscs}
PACS-CS Collaboration, S.~Aoki {\it et al.},
{\tt arXiv:0807.1661};\\
\bibitem{kura_lat08}
Y.~Kuramashi,
\pos{PoS(LATTICE 2008)018}.
\bibitem{ukita_lat08}
PACS-CS Collaboration, N.~Ukita {\it et al.},
\pos{PoS(LATTICE 2008)097}.
\bibitem{kadoh_lat08}
PACS-CS Collaboration, D.~Kadoh {\it et al.},
\pos{PoS(LATTICE 2008)092}.\\


\bibitem{param} 
S.~Aoki, Y.~Kayaba and Y.~Kuramashi,
{\em Nucl. Phys. B} {\bf 697} (2004) 271.


\bibitem{PDG}
C. Amsler {\it et al.},
{\em Phys. Let. B} {\bf 667} (2008) 1.



\bibitem{hfs_hpqcd}
E.~Follana {\it et al.},
{\em Phys. Rev.} {\bf D75}, 054502 (2007).

\bibitem{kura_lat05}
CP-PACS Collaboration, Y.~Kuramashi {\it et al.},
\pos{PoS(LATTICE 2005)226}.

\bibitem{RHQ-N_f_0_2}
CP-PACS Collaboration, Y.~Kayaba {\it et al.},
{\em JHEP} {\bf 0702} (2007) 019;
{\em Nucl.Phys.Proc.Suppl.} {\bf 140} (2005) 479.


\bibitem{zvza}
S.~Aoki, Y.~Kayaba and Y.~Kuramashi,
{\em Nucl. Phys. B} {\bf 689} (2004) 127.


\bibitem{NP-c_A}
CP-PACS/JLQCD and ALPHA Collaboration, T.~Kaneko {\it et al.},
{\em JHEP} {\bf 0704} (2007) 092.


\bibitem{CLEO}
S.~Stone, {\tt arXiv:0806.3921}.

\bibitem{hpqcd}
HPQCD and UKQCD Collaborations, E.~Follana {\it et al.},
{\em Phys. Rev. Lett.} {\bf 100} (2008) 062002.


\bibitem{fnal}
FNAL Lattice, HPQCD and MILC Collaborations, C.~Aubin {\it et al.},
{\em Phys. Rev. Lett.} {\bf 95} (2005) 122002.

\end{thebibliography}
\end{document}